\documentclass[10pt,a4paper,preprint,english,aip,jcp,preprint,superscriptaddress,onecolumn]{revtex4-1}
\usepackage[T1]{fontenc}
\usepackage[latin9]{inputenc}
\usepackage{color}
\usepackage{verbatim}
\usepackage{amsmath}
\usepackage{graphicx}
\usepackage{amssymb}
\usepackage{esint}

\makeatletter

 \@ifundefined{textcolor}{}
 {%
   \definecolor{BLACK}{gray}{0}
   \definecolor{WHITE}{gray}{1}
   \definecolor{RED}{rgb}{1,0,0}
   \definecolor{GREEN}{rgb}{0,1,0}
   \definecolor{BLUE}{rgb}{0,0,1}
   \definecolor{CYAN}{cmyk}{1,0,0,0}
   \definecolor{MAGENTA}{cmyk}{0,1,0,0}
   \definecolor{YELLOW}{cmyk}{0,0,1,0}
 }

\@ifundefined{definecolor}
 {\usepackage{color}}{}

\newif\iftwocolumn
\twocolumntrue

\usepackage{graphicx}
\usepackage{dcolumn}
\usepackage{subfigure}
\usepackage{multirow}
\usepackage{bm}

\newcommand{\ab}[1]{\mathrm{#1}}

\newcommand{\onetep}{{\sc onetep}}
\newcommand{\amber}{{\sc amber}}

\newcommand{\etal}{\textit{et al.}}
\newcommand{\abinitio}{\textit{ab initio}}

\newcommand{\bvec}[1]{\textbf{#1}}

\newcommand{\eps}{\varepsilon}
\newcommand{\ofr}{\left(\bvec{r}\right)}
\newcommand{\epsofr}{\eps\ofr}
\newcommand{\epsofrho}{\eps\left[\rho\right]}
\newcommand{\rhoofr}{\rho\ofr}
\newcommand{\rhotot}{\rho_{\ab{tot}}}

\newcommand{\oldrevision}[1]{#1}

\newcommand{\revision}[1]{#1}

\makeatletter
\newlength \figwidth
\iftwocolumn
\else
\fi
\newlength \thirdofpage
\makeatother

\begin{document}

\title{Minimal parameter implicit solvent model for \abinitio{} electronic structure calculations}

\author{J. Dziedzic}
\altaffiliation{Also at: Faculty of~Technical Physics and Applied Mathematics, Gdansk University of~Technology, Narutowicza~11/12, 80-952 Gda\'{n}sk, Poland}

\affiliation{
  School of Chemistry, University of Southampton, Highfield, Southampton SO17 1BJ, United Kingdom
}

\author{H. H. Helal}

\affiliation{
  Theory of Condensed Matter group, Cavendish Laboratory, University of Cambridge, Cambridge CB3 0HE, United Kingdom
}

\author{C.-K. Skylaris}

\affiliation{
  School of Chemistry, University of Southampton, Highfield, Southampton SO17 1BJ, United Kingdom
}

\author{A. A. Mostofi}

\affiliation{
  The Thomas Young Centre for Theory and Simulation of Materials, Imperial College London,  London SW7 2AZ, United Kingdom
}

\author{M. C. Payne}

\affiliation{
  Theory of Condensed Matter group, Cavendish Laboratory, University of Cambridge, Cambridge CB3 0HE, United Kingdom
}

\begin{abstract}
We present an implicit solvent model for \abinitio{} electronic structure calculations which
is fully self-consistent and is based on direct solution of the nonhomogeneous Poisson equation.
The solute cavity is naturally defined in terms of an isosurface of the electronic density
according to the formula of Fattebert and Gygi (J.~Comp.~Chem.~\textbf{23}, 6 (2002)).
While this model depends on only two parameters, we demonstrate that by using appropriate
boundary conditions and dispersion-repulsion contributions, solvation energies obtained for an extensive test set including neutral and charged molecules show dramatic improvement compared to existing models.
Our approach is implemented in, but not restricted to, a linear-scaling density functional theory (DFT) framework, opening the path for self-consistent implicit solvent DFT calculations on systems of unprecedented size, which we demonstrate with calculations on a 2615-atom protein-ligand complex. 
\end{abstract}

\maketitle

The role of solvent is critical to a multitude of chemical, biological and physical processes\oldrevision{.} The accurate simulation of such processes, therefore, requires careful treatment of solvation effects.
However, explicit \mbox{inclusion} of the solvent with 
full atomic detail \oldrevision{is} very costly due to the significant increase in the number
of simulated atoms and the need for extensive averaging over~the solvent degrees of freedom~\cite{henchman1999}. Moreover, such explicit treatment may also be unnecessary, as it is often the long-range electrostatic effect of the solvent that is most significant, with only a small
proportion of solvent molecules \oldrevision{involved chemically}. 
The implicit solvent approach addresses these issues by retaining only the ato\-mic details of the solute, placed in a suitably defined cavity, and by representing the solvent environment by an unstructured dielectric continuum outside this cavity.
The free energy of solvation is typically decomposed into two contributions -- the electrostatic energy of interaction of the solvent with the polarized dielectric, and a nonpolar term accounting for the work required to create a cavity in the solvent (cavitation energy), and, in more \oldrevision{complex} models, for dispersion-repulsion interactions between the solute and solvent.  


A multitude of 
implicit solvent models of differing sophistication
have been proposed to date~\cite{Tomasi2} for use in \abinitio{} calculations. 
Many of these models are derived from the self-consistent reaction field (SCRF)
formalism, where the effect of the electric field due to the dielectric (polarized by the solute) is
included in the Hamiltonian in a self-consistent fashion.   
Two widely used classes of SCRF-type models are the polarizable continuum 
model (PCM) of Tomasi \etal{} \cite{Tomasi1}
and the conductor-like screening model (COSMO) of Klamt and Sch\"{u}\"{u}rmann \cite{Klamt1993}.

The shape of the cavity containing the solute varies between models -- early models used spherical or \mbox{elliptical} cavities; in more recent models the cavity is usually constructed from overlapping atomic spheres of varying ra\-di\-i, which necessitates using a number of parameters. In contrast, the recent model
proposed by Fattebert and Gygi \cite{FattebertGygi,FattebertGygi2003}, and later developed by Scherlis \etal{} \cite{Scherlis} (henceforth called the FGS model), utilizes a dielectric cavity constructed directly from the electronic density of the~solute, which greatly reduces the number of parameters involved.

This ``minimal-parameter'' nature of the FGS model makes it attractive for \abinitio{} calculations. However, this approach also has several shortcomings, which we address in this Letter.
First, the \oldrevision{original model did not} include dis\-per\-sion-repulsion effects, which, as the authors themselves note, is likely to 
impact its accuracy for larger neutral molecules. Second, \oldrevision{%
the model necessitates the use of an \textit{a posteriori} correction to the energy in vacuum,
obtained in periodic boundary conditions, to approximate open boundary conditions, whereas \revision{in the} solvent the electrostatic energy is obtained subject to zero boundary conditions.}
Third, a severe numerical instability prevents this approach from being practical for large molecules.

This Letter describes how we have addressed these limitations, by including dispersion-repulsion interactions, employing \oldrevision{open (Coulombic)} boundary conditions, and identifying and circumventing the root cause of the abovementioned numerical instability. We then validate and evaluate the performance of the model on two sets of several tens of small molecules. Finally, by per\-for\-ming a calculation on a 2615-atom protein-ligand system, we demonstrate how the implemented model can be used to perform large-scale \abinitio{} calculations in solution. 

In contrast to other SCRF models where the solute cavity has a discontinuous boundary, the FGS model defines a smooth transition of the \oldrevision{relative} permittivity according to:
\begin{equation}
  \epsofr = 1+\frac{\varepsilon_\infty-1}{2}
  \left(1+\frac{1-{\left(\rhoofr/\rho_0\right)}^{2\beta}}{1+{\left(\rhoofr/\rho_0\right)}^{2\beta}}\right),
\label{eq:eps}
\end{equation}
where $\rhoofr$ is the electronic density of the solute, $\varepsilon_\infty{}$ is the bulk permittivity, the parameter $\beta$ 
controls the smoothness of the transition of $\epsofr$ from 1 to $\varepsilon_\infty$,  
and $\rho_0$ is the density value for which the permittivity drops to $\varepsilon_\infty{}/2$. 
The cavitation contribution to the free energy is assumed to be proportional to the surface area, $S$, of the cavity (calculated at $\rho=\rho_0$), that is $\Delta{}G_{\ab{cav}}=\gamma S(\rho_0)$, where $\gamma$ is the solvent surface tension. 
Values for $\beta$ and $\rho_0$ are found by a least-squares fit to the hydration energies of ammonia, nitrate and methylammonium (representative of neutral, anionic and cationic molecules, respectively) 
\cite{Scherlis}.  
The total potential of the solute in 
the presence of the dielectric, $\phi\ofr$ is obtained by solving
the nonhomogeneous Poisson equation 
\begin{equation}
  \nabla \cdot \left(\epsofrho \nabla \phi\right) = -4\pi\rhotot
\label{eq:NPE}
\end{equation}
directly in real space subject to zero 
Dirichlet boundary conditions. The total charge density $\rhotot\ofr$ is a sum of the electronic density $\rhoofr$ and a Gaussian-smeared density of the cores, as proposed in ref.~\onlinecite{Scherlis}. 

As outlined in ref.~\onlinecite{Scherlis}, the fact that the dielectric cavity responds self-consistently to changes in the electronic density means that the functional derivative of the electrostatic energy, $E_{\ab{es}}$, is no longer equal to the potential that is the solution of eq.~(\ref{eq:NPE}), \oldrevision{but rather:}
\oldrevision{
\begin{equation}
  \frac{\delta{}E_{\ab{es}}}{\delta\rho}\ofr=\phi\ofr-\frac{1}{8\pi}{\left(\nabla\phi\ofr\right)}^2\frac{\delta\eps}{\delta\rho}\ofr.
\label{eq:dielcorr}
\end{equation}
}
\begin{figure}[htb!]
\includegraphics[viewport=10 100 505 635,angle=-90,clip,width=\thirdofpage]{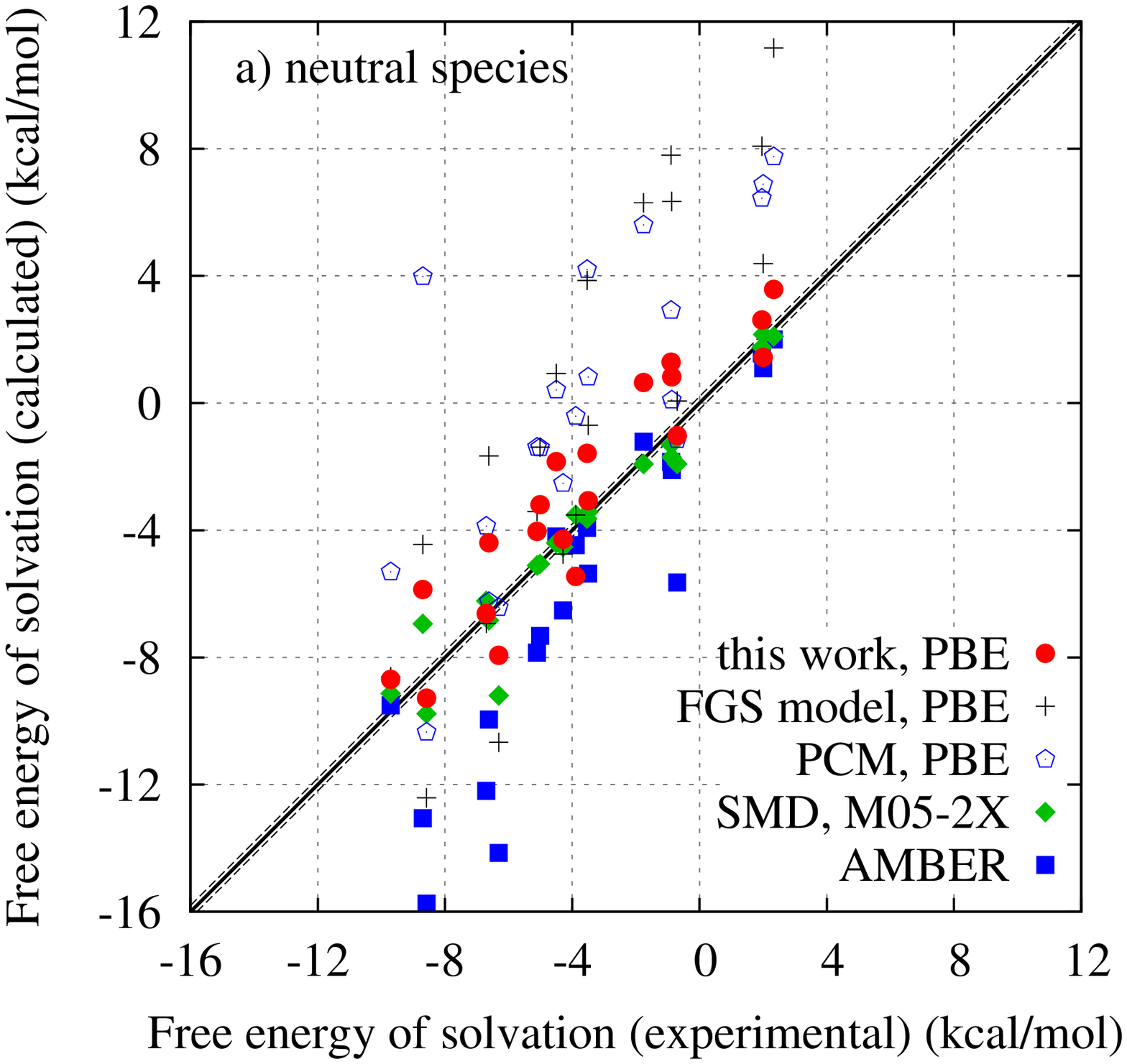}
\includegraphics[viewport=10 100 505 635,angle=-90,clip,width=\thirdofpage]{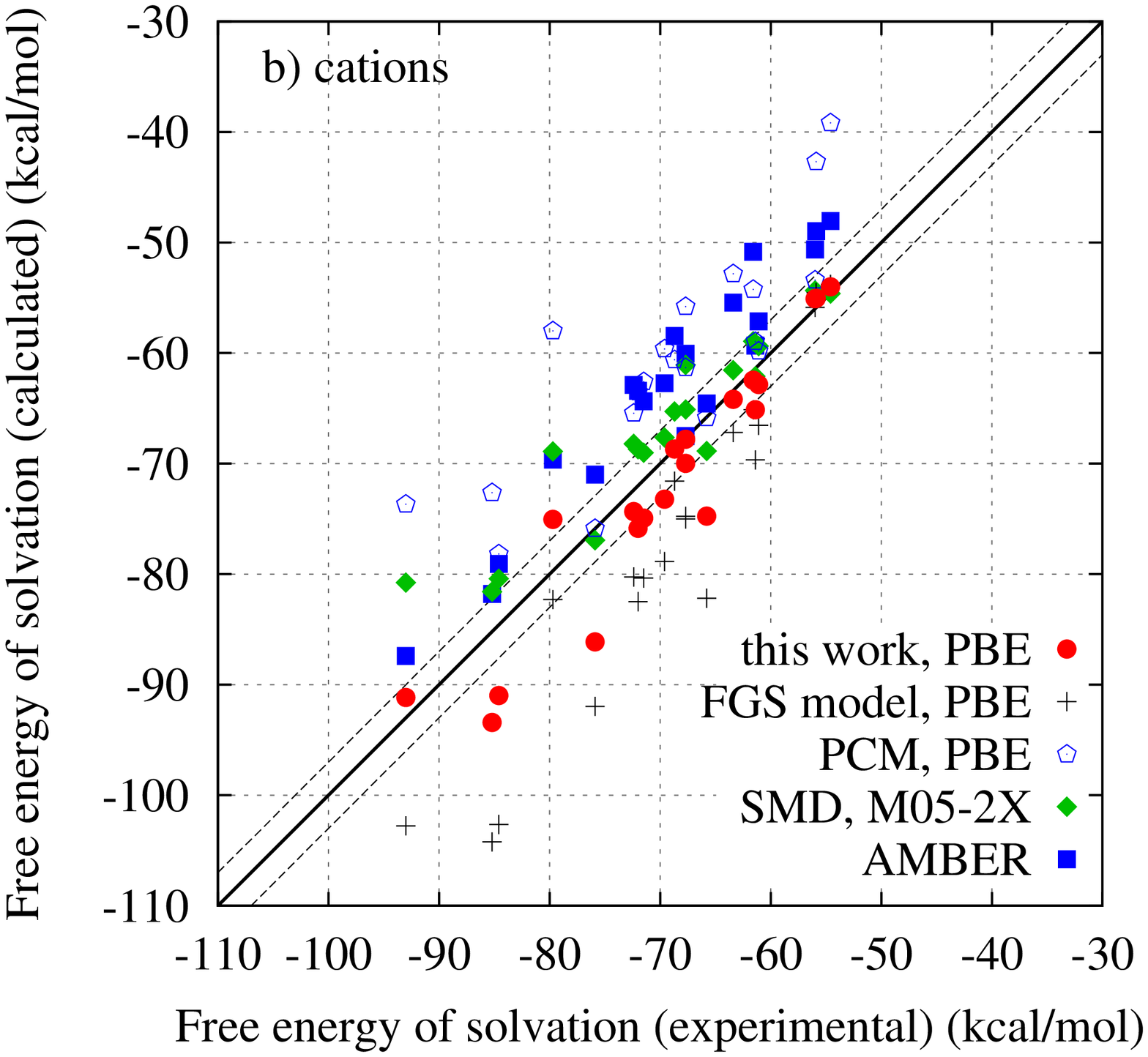}
\includegraphics[viewport=10 100 505 635,angle=-90,clip,width=\thirdofpage]{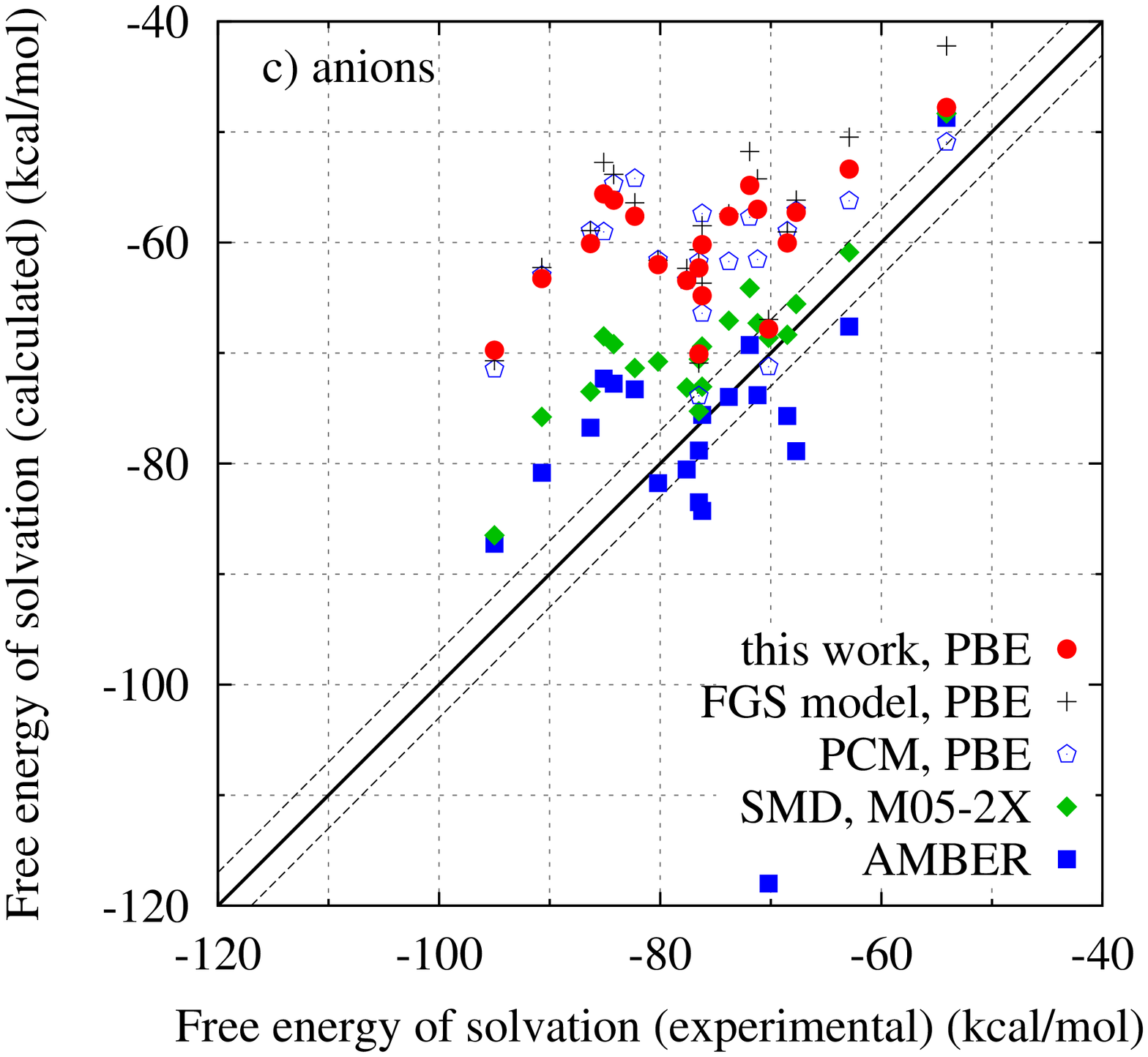}
\caption{
\label{fig:NAC60_onetep_vs_others} 
Calculated free energies of solvation plotted against corresponding experimental values -- a comparison between this work and other models.   
The solid diagonal line represents perfect agreement with experiment, the dashed black lines denote the estimated uncertainty of the experimental values.  
}
\end{figure}

The \oldrevision{original} FGS model does not set out to address dispersion-repulsion effects. This makes the results obtained for larger molecules dubious, especially for those that are neutral, as then the electrostatic contribution to solvation would be dwarfed by the nonpolar terms. As the authors duly note, this deficiency already becomes evident for the case of benzene where this model predicts a $\Delta{}G$ of 7.9 kcal/mol \cite{Scherlis} whereas the experimental value is \mbox{-0.87}~kcal/mol \cite{soldb}. To appreciate the magnitude of the problem we refer to fig.~\ref{fig:NAC60_onetep_vs_others} and the top of table~\ref{tab:main}, where results obtained with the FGS model are shown for a representative selection of 20 neutral, 20 cationic and 20 anionic molecules chosen from ref.~\onlinecite{soldb}. The geometries
of the molecules were not re-optimized in solvent, instead geometries optimized in the gas phase readily available from ref.~\onlinecite{soldb} were used. \oldrevision{The FGS model underestimates the solvation effect for anions and overestimates it for cations. The predictions for neutral species are in
moderate agreement with experiment.}
By examining the coefficient of correlation between the calculated and experimental values, we demonstrate that the obtained values do not correlate well with experiment (with the notable exception of cations), which makes the calculation of \textit{relative} free energies of solvation, $\Delta\Delta{}G$, unreliable.

The second shortcoming of this approach
is related to the boundary conditions used for the solution of eq.~(\ref{eq:NPE}). References \cite{FattebertGygi,FattebertGygi2003} are concerned only with calculations in solution, where 
zero boundary conditions are used. Owing to the dielectric screening, this is a reasonable
approximation, as long as the relative permittivity of the solvent is large. 
Ref.~\cite{Scherlis} uses the same approach in
solution, whereas for the reference vacuum calculation (needed to obtain free energies of solvation),
where the Poisson equation becomes homogeneous, standard periodic plane-wave DFT calculations are performed. \textit{In vacuo} energies thus obtained are subsequently corrected with the
Makov-Payne formula \cite{MakovPayne} to mimic the effect of open boundary conditions. 
This too is an approximation, since the correction cannot fully capture polarization effects \cite{MakovPayne}.
Furthermore, only the energy is corrected, while the shape of the electronic density, and, in
turn, the cavity generated in solution corresponds to periodic boundary conditions. As we demonstrate later, this subtly affects the free energies of solvation obtained for charged species, leading to a degree of cancellation of errors.

Further, we point out the root cause of the numerical instability inherent in the FGS model. The second term in the RHS of eq.~(\ref{eq:dielcorr}) is extremely difficult to evaluate accurately, because $\frac{\delta\eps}{\delta\rho}$ is very close to zero everywhere, except on the boundary of the cavity, where, in turn, ${\left(\nabla\phi\ofr\right)}^2$ is almost zero and thus difficult to distinguish from numerical noise. Because of this, the energy gradient calculated from eq.~(\ref{eq:dielcorr}) is not numerically accurate and the method is found to converge only when high-order finite-differences and extremely fine grids (with a spacing of $0.15\,a_0$ or finer) are used, as only then the gradient of the potential can be evaluated to sufficient accuracy. The memory requirements necessitated by such fine grids quickly make the technique impractical for larger molecules.

\begingroup
\begin{table}
\caption{
\label{tab:main}
Error (root mean square (rms) and maximum, in kcal/mol), with respect to experiment \cite{soldb}, in the calculated free energies of solvation, and the corresponding coefficient of correlation, $r$, between the calculated and experimental values, for the 20 neutral, 20 cationic and 20 anionic species studied.
}
\begin{tabular}{ll|rrr|rrr|rrr}
 &XC&\multicolumn{3}{c|}{neutral species}&\multicolumn{3}{c|}{cations}&\multicolumn{3}{c}{anions}\\
 Approach&functional&
 rms err.&max err.&$r$&
 rms err.&max err.&$r$&
 rms err.&max err.&$r$
\\ \hline
 \oldrevision{FGS}&\oldrevision{PBE}&\oldrevision{5.0}&\oldrevision{8.8}&\oldrevision{0.87}&\oldrevision{9.7}&\oldrevision{19.0}&\oldrevision{0.95}&\oldrevision{19.5}&\oldrevision{32.4}&\oldrevision{0.55} \\
 this work\footnote{With the cavity responding self-consistently to changes in density.}
&PBE&1.6&2.8&0.93&4.4&10.2&0.95&18.1&29.5&0.53 \\
 \oldrevision{this work}\footnote{\oldrevision{With the cavity responding self-consistently to changes in density, and without dispersion-repulsion.}}&\oldrevision{PBE}&\oldrevision{5.0}&\oldrevision{8.9}&\oldrevision{0.87}&\oldrevision{10.4}&\oldrevision{19.0}&\oldrevision{0.95}&\oldrevision{21.2}&\oldrevision{35.1}&\oldrevision{0.54} \\  this work\footnote{With the cavity fixed.}&PBE&1.8&3.1&0.93&3.9&8.3&0.94&18.1&29.4&0.54 \\
PCM&PBE&4.9&12.7&0.75&10.5&21.7&0.83&17.8&29.5&0.36 \\
 PCM&B3LYP&4.7&12.0&0.78&10.4&21.8&0.83&17.0&28.4&0.41 \\
 PCM&M05-2X&4.4&11.1&0.79&10.2&21.7&0.81&15.7&26.8&0.46 \\
 SMD&M05-2X&0.9&2.9&0.97&4.6&12.2&0.95&8.5&16.6&0.86 \\
 \amber{} \cite{amber}&(classical)&3.3&7.84&0.64&6.9&10.8&0.96&12.8&47.8&0.32 \\
\end{tabular}
\end{table}
\endgroup

By addressing each of these limitations, we obtain a highly accurate and
usable approach which retains the conceptual elegance of the FGS model. 

\oldrevision{We solve eq.~(\ref{eq:NPE}) by means of a second-order multigrid\cite{multigrid,Holst} approach, which is subsequently defect-corrected\cite{defcorr} in an iterative fashion using 10-th order finite-difference stencils for the first and second derivatives. We find that with a grid spacing of $0.125\,a_0$ as few as 3-4 defect-correction iterations are sufficient to reduce the algebraic
error in the obtained potential by four orders of magnitude with respect to the initial, uncorrected
solution. The corresponding reduction in the magnitude of the residual is two orders of magnitude,
due to the approximate nature of the calculated boundary conditions. With a grid spacing of $0.25\,a_0$
ten iterations, on average, were necessary.}

\begin{figure}[htbp!]
\includegraphics[viewport=25 170 490 570,width=0.54\figwidth,
angle=-90,clip]{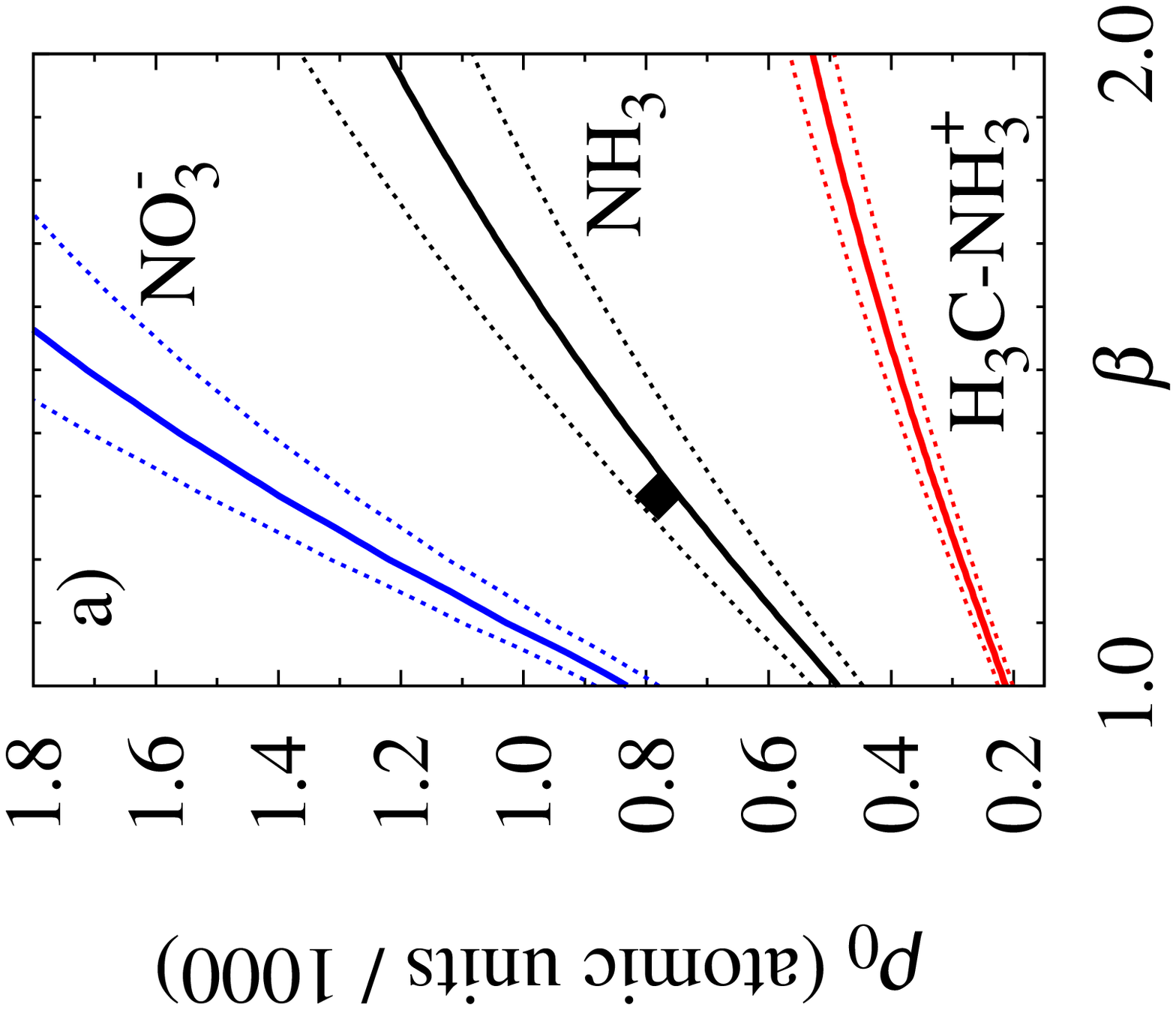}
\includegraphics[viewport=25 265 490 555,width=0.54\figwidth,
angle=-90,clip]{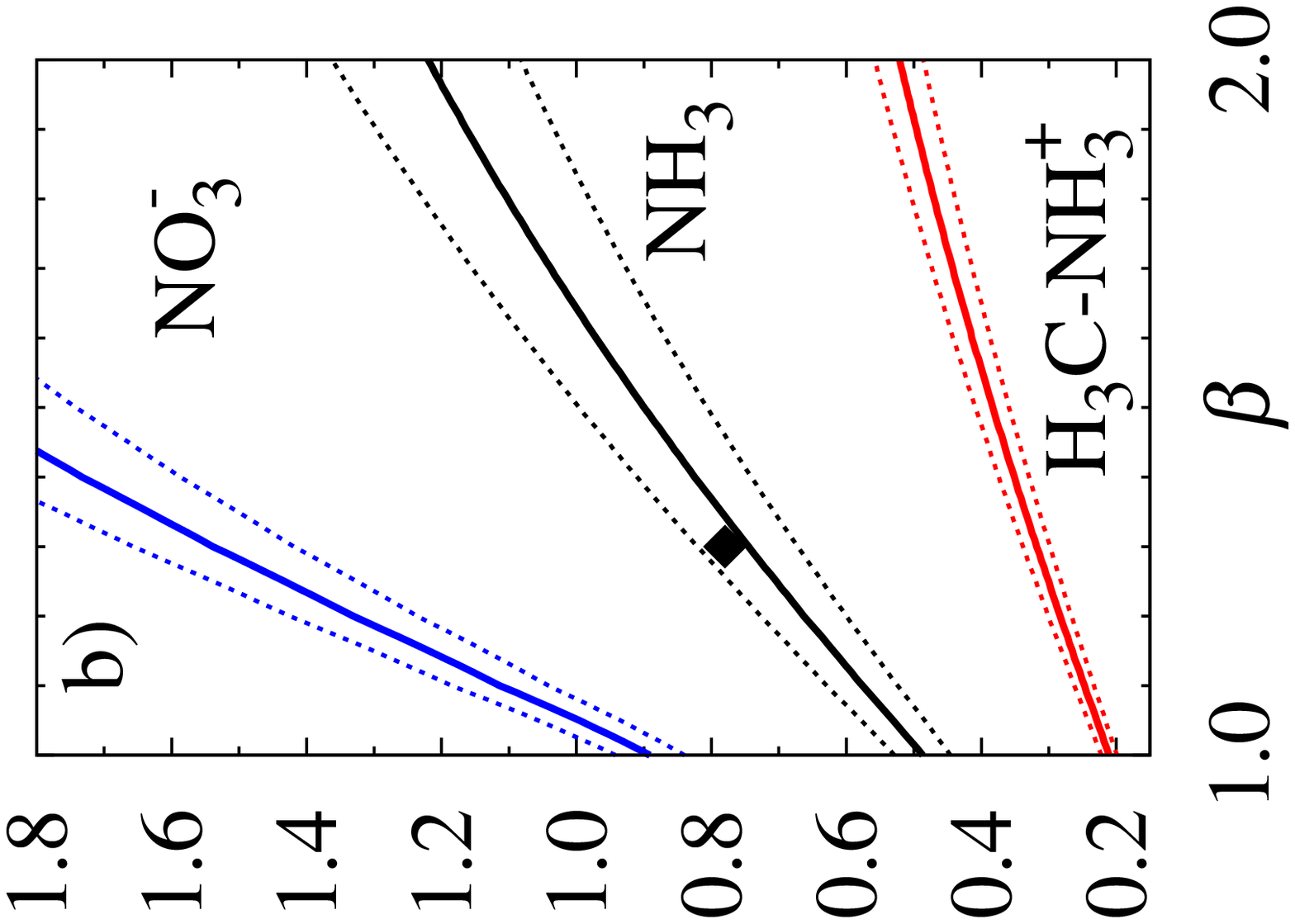}
\includegraphics[viewport=25 265 490 555,width=0.54\figwidth,
angle=-90,clip]{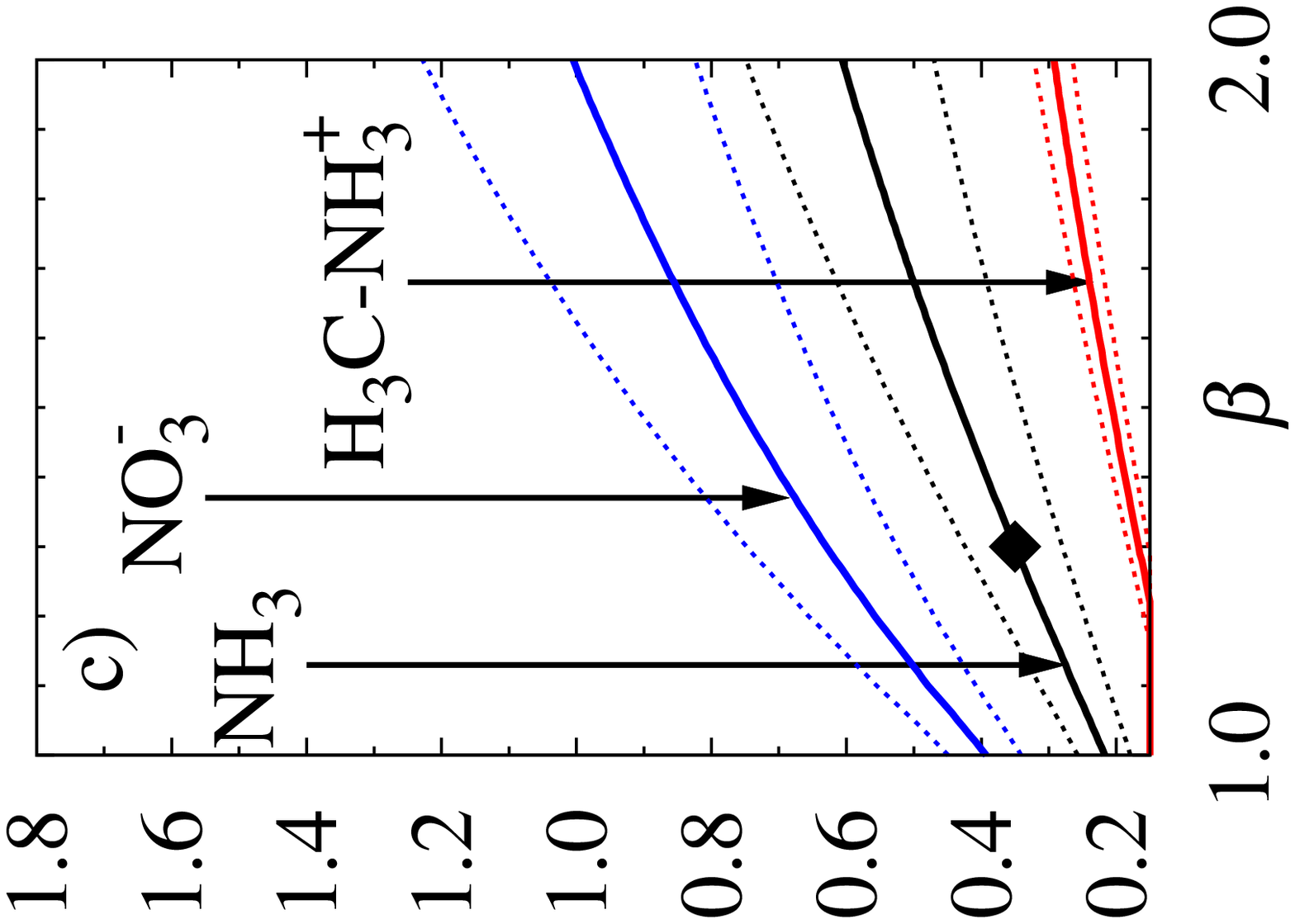}
\caption{
\label{fig:sweep} 
Isolines of zero error with respect to experiment for the possible parametrizations of a) the FGS model,
b) the FGS model with \oldrevision{the boundary conditions we propose} and c) the model proposed in this work, for three
representative molecules. The dashed lines indicate an error of $\pm 1$~kcal/mol. 
The black points indicate the final parametrization of \oldrevision{the respective models.}
}
\end{figure}

\begin{figure*}[ht!]
\includegraphics[viewport=10 100 505 635,angle=-90,clip,width=\thirdofpage]{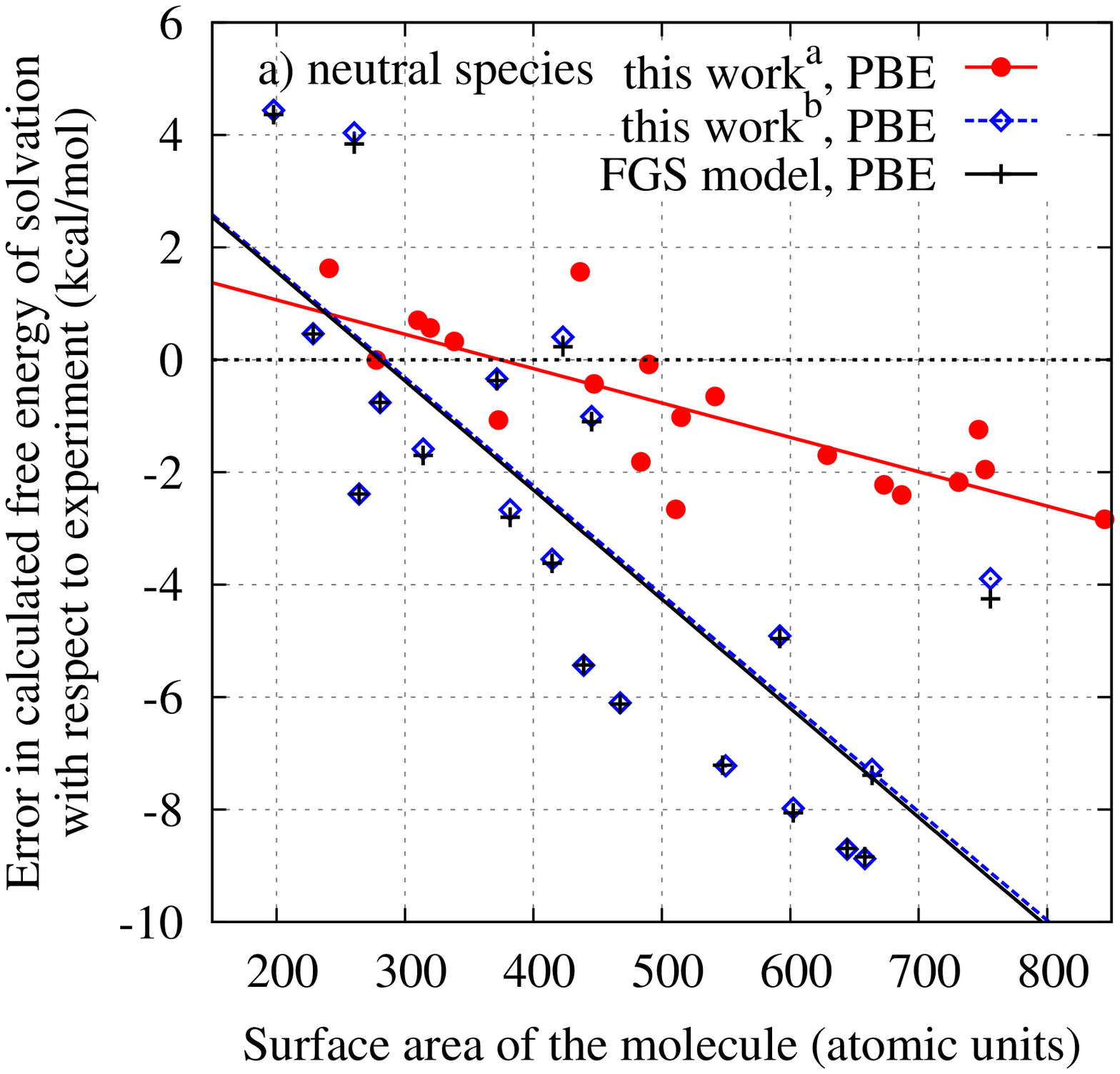}
\includegraphics[viewport=10 100 505 635,angle=-90,clip,width=\thirdofpage]{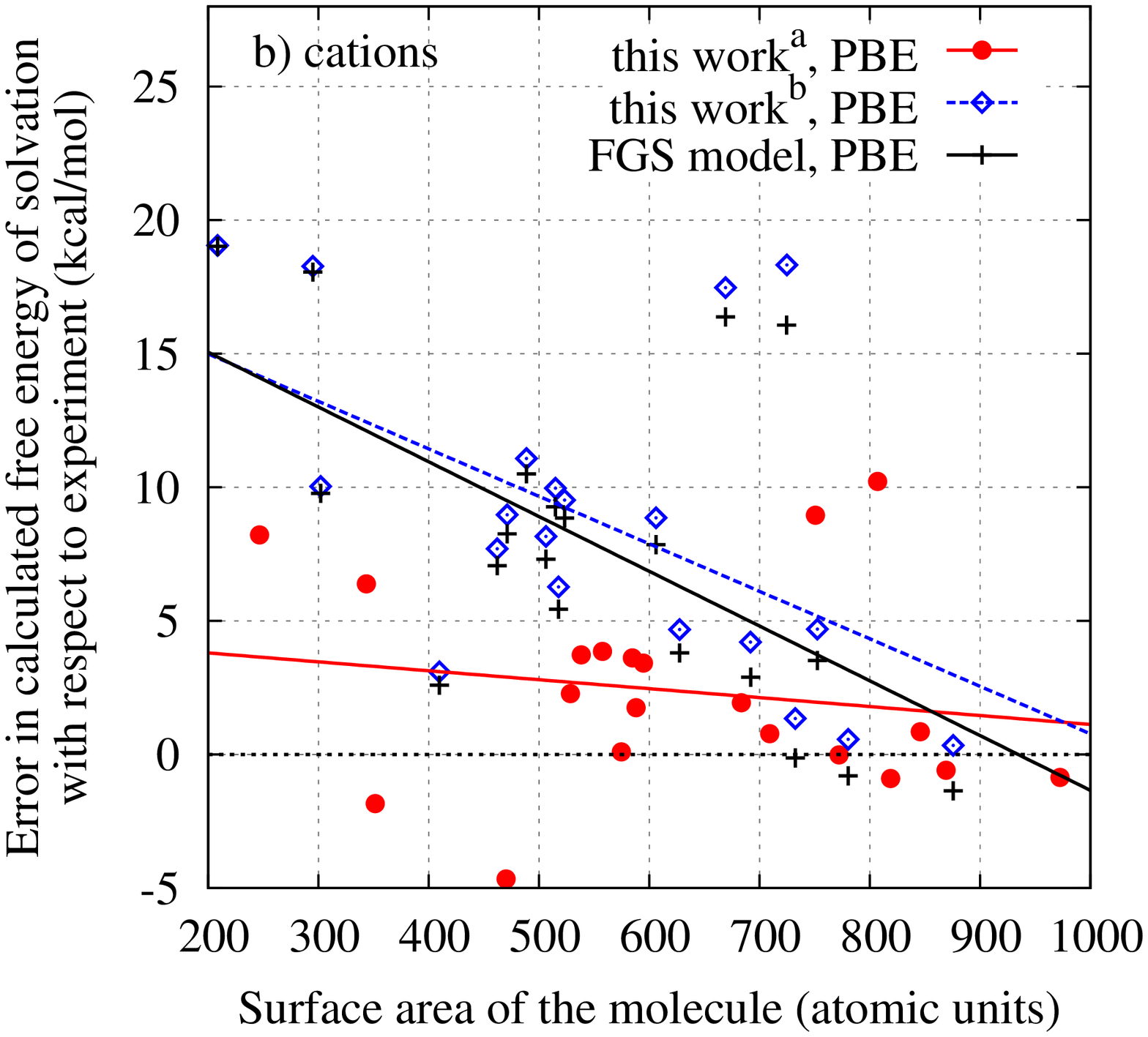}
\includegraphics[viewport=10 100 505 635,angle=-90,clip,width=\thirdofpage]{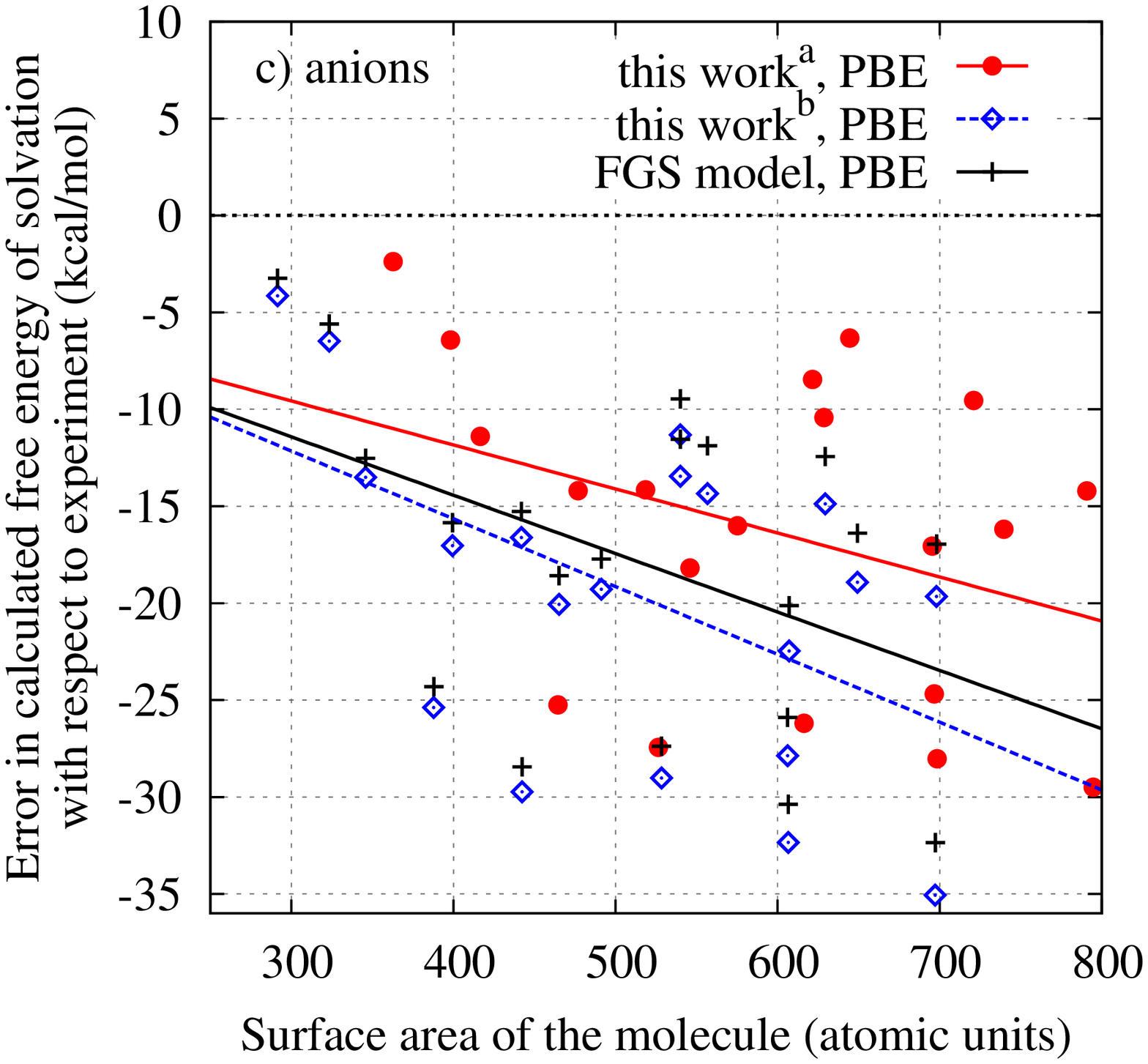}
\caption{
\label{fig:NAC60_error_vs_surfarea} 
Absolute error in the calculated free energies of solvation with respect to experimental values, plotted
against the surface area of the molecule. The surface areas differ between models because the
parametrization is different. The solid and dashed lines represent a linear fit. The horizontal dotted
line corresponds to perfect agreement with experiment.
}
\end{figure*}

We have recast the solvation problem into open boundary conditions by computing the core-core
and the local pseudopotential terms in real space and by using \oldrevision{open (Coulombic)} boundary conditions when solving eq.~(\ref{eq:NPE}) -- that is, we set up Dirichlet boundary conditions by evaluating the Coulombic potential due to the charge distribution $\rhotot$. Since with a spatially localized density the calculation of the Coulombic integral for all the points on the boundary scales as $\mathcal{O}\left(L^2 N_{\ab{at}}\right)$ with the box length $L$ and number of atoms $N_{\ab{at}}$, charge coarse-graining and interpolation were used to reduce the prefactor in this calculation by about three orders of magnitude. \oldrevision{In so doing, we obtain \textit{in vacuo} energies and densities
that need not be corrected.}  
In the solvated case, where the nonhomogeneity in $\eps$ prevents such an approach, \oldrevision{we calculate} the boundary conditions by approximating the dielectric as homogeneous, with the permittivity of the bulk solvent. \oldrevision{Fig.~\ref{fig:sweep}, panels a) and b) and table~\ref{tab:main}, rows 1 and 3 illustrate that for the simulation cell we used (a cube with an edge length of $L=40.5\,a_0$) this alone offers no improvement in accuracy. A difference in the free energy of solvation of 0.1~kcal/mol, 0.9~kcal/mol and -1.8~kcal/mol is observed on average for neutral, cationic and anionic species, respectively, compared to the predictions of the original FGS model (refer to fig.~\ref{fig:NAC60_error_vs_surfarea} for details). For charged species the application of consistent open boundary conditions leads to a slight increase in the error.}

A parameter sweep for the three molecules used to parametrize the FGS model demonstrates (cf.~fig.~\ref{fig:sweep}) that there exists no parametrization that would result in even moderate agreement with experiment for the three species simultaneously. The model would consistently either underestimate free energies of solvation for anions or overestimate \oldrevision{them} for cationic species. 

We attribute this failure to a combination of factors -- the poor performance of the Perdew-Burke-Ernzerhof (PBE) exchange-correlation (XC) functional; the fact that \textit{any} isodensity formulation will use larger cavities for anions than for corresponding cations, whereas the charge assymmetry in solvation effects is in fact opposite \cite{Mobley_asymmetry}; and, finally, to the lack of inclusion of dispersion-repulsion effects, which leads to an overestimation of the nonpolar component of solvation. The middle rows of table~\ref{tab:main} show, on the example of PCM, how using a hybrid functional such as B3LYP or a hybrid meta-GGA functional such as M05-2X \cite{M052X} addresses the first problem, by reducing the self-interaction error, which otherwise leads to excessive delocalization of the electrons, but does not address the other two problems. 

\oldrevision{The increase in the magnitude
of the difference between the calculated and experimentally obtained free energies of solvation
with the size of the molecule, especially in the case of neutral molecules, demonstrated in fig.~\ref{fig:NAC60_error_vs_surfarea} indicates that the neglect of dispersion-repulsion effects
is detrimental to the predictive quality of the FGS model.} 
We propose including dispersion-repulsion effects in the free energy of solvation, $\Delta{}G_{\ab{dis,rep}}$, using an approximate relation derived by Floris \etal{} \cite{FlorisTomasi}. Since~this relation is linear, it amounts to a simple rescaling of the surface tension of the solvent, including the approximate $\Delta{}G_{\ab{dis,rep}}$ in the cavitation term. \revision{From the slope
of the linear relation plotted in fig.~1 of ref.~\cite{FlorisTomasi} it follows that the
surface tension should be rescaled by a factor of 0.281.} Even this crude method for taking dispersion-repulsion into account dramatically improves the accuracy of the model, as evidenced by figs.~\ref{fig:NAC60_onetep_vs_others} and \ref{fig:sweep} and table~\ref{tab:main}, from which it is apparent that the resulting approach is in much better agreement with experiment than both PCM and the force-field Poisson-Boltzmann (PB) approach of \amber{} \cite{amber}, offering comparable quality to the much more complex SMD\footnote{
SMD is a recently proposed model based on the integral-equation-formalism PCM (IEF-PCM), which yields excellent agreement with experiment. This requires, however, making use of an extensive set of parameters to describe the solute (intrinsic Coulomb radii, atomic surface tension parameters) and the solvent (refractive index and acidity and basicity parameters; in addition to the dielectric constant and bulk surface tension needed in the proposed model). 
}~\cite{SMD} model. \oldrevision{The improvement offered by the inclusion of dispersion-repulsion effects
is evidenced by fig.~\ref{fig:NAC60_error_vs_surfarea} and can be quantified by comparing rows denoted with $a$ and $b$ in table~\ref{tab:main}. The results corresponding to the row denoted with $b$ were obtained by turning off the dispersion-repulsion contribution whilst using the parameters proposed in ref.~\cite{Scherlis}, denoted with a point in fig.~\ref{fig:sweep}, panels a) and b).}  

The numerical instability caused by \oldrevision{the second term} in the RHS of eq.~(\ref{eq:dielcorr}) can be circumvented without loss of accuracy. We first note that \oldrevision{this term disappears} when, instead of responding to changes in the electronic density, the dielectric cavity is fixed. We propose constructing the cavity by the application of eq.~(\ref{eq:eps}) to the converged electronic density of the solute obtained in the vacuum calculation and keeping the cavity fixed throughout the calculation in solvent. We show (cf.~tables~\ref{tab:main},~\ref{tab:blind_test}) that the associated reduction in accuracy is insignificant, while both the wall time and the memory requirements of the computation are reduced by about an order of magnitude, as convergence is readily achieved with a more reasonable 
real-space grid spacing of $0.25\,a_0$. \oldrevision{We should point out that a similar attempt to fix the cavity 
in the FGS model would probably lead to larger errors due to the fact that the fixed cavity would come from the \textit{periodic} density of the vacuum calculation -- as the Makov-Payne correction \cite{MakovPayne} only corrects the energy.} We note that this simplified approach is still
suitable for geometry optimization in solution, provided the additional contribution to the forces due to the cavity variation with atomic positions is included. S\'{anchez} \etal{} \cite{Scherlis2009} also note the abovementioned instability and propose a somewhat different way of circumventing it. 
\begin{table}
\begin{minipage}{\columnwidth}
\centering
\caption{\label{tab:blind_test}
Error (root mean square (rms) and maximum, in kcal/mol), with respect to experiment, in the calculated free energies of solvation, and the corresponding coefficient of correlation, $r$, between calculated and experimental values for the 71 molecules studied \cite{blind1,blind2}.
}
\begin{tabular}{ll|rrr}
  &&rms&max&\\  
  Approach&XC functional&error&error&$r$
\\ \hline
 this work\footnote{With cavity responding self-consistently to changes in density.}&PBE&3.8&8.3&0.83 \\
 this work\footnote{With cavity fixed.}&PBE&4.1&9.1&0.83 \\
 PCM&PBE&10.9&23.3&0.53 \\
 SMD&M05-2X&3.4&14.5&0.87 \\
 \amber{}&(classical)&5.1&19.9&0.77 \\
\end{tabular}
\end{minipage}
\end{table}
\begin{figure}[htbp]
\centering
\includegraphics[viewport=10 110 505 620,
angle=-90,clip,width=\figwidth]{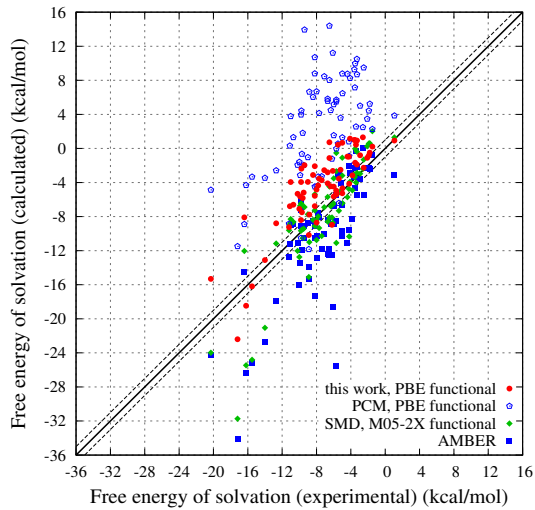}
\caption{
\label{fig:test71_onetep_vs_others} 
Calculated free energies of solvation plotted against corresponding experimental values for the 71 neutral molecules studied \cite{blind1,blind2} -- a comparison between models. 
The solid diagonal line represents perfect agreement with experiment, the dashed black lines denote the estimated uncertainty in the experimental valus.  
}
\end{figure}

We further validate our model on 71 neutral molecules taken from the blind tests of refs.~\onlinecite{blind1,blind2}, for which the experimental energies of solvation are reported in  ref.~\cite{soldb}. Again, the geometries were not re-optimized in solution, but rather the 
gas-phase geometries from ref.~\cite{soldb} were used. The results, shown in fig.~\ref{fig:test71_onetep_vs_others} and table~\ref{tab:blind_test}, again show that our approach 
is consistently more accurate than both PCM and the force-field PB approach of \amber{} \cite{amber} and that our model offers a level of agreement with experiment which is comparable to the SMD \cite{SMD} model, even when the cavity is fixed.

\begin{table}
\caption{\label{tab:lysozyme}
Free energies of solvation (in kcal/mol) of L99A/M102Q T4 lysozyme ($\Delta{}G_{\ab{host}}$), its complex with catechol ($\Delta{}G_{\ab{cplx}}$), catechol ($\Delta{}G_{\ab{lig}}$), desolvation energy of the ligand ($\Delta{}G_{\ab{d}}=\Delta{}G_{\ab{cplx}}-\Delta{}G_{\ab{host}}-\Delta{}G_{\ab{lig}}$), binding energy \textit{in vacuo} ($\Delta{}E_{\ab{gas}}$) and in solvent ($\Delta{}E_{\ab{sol}}$), as predicted by our model (with PBE and a fixed cavity) and \amber{}.
}
\begin{tabular}{l|rrrrrrr}
Approach&$\Delta{}G_{\ab{cplx}}$&$\Delta{}G_{\ab{host}}$&$\Delta{}G_{\ab{lig}}$&$\Delta{}G_{\ab{d}}$&$\Delta{}E_{\ab{gas}}$&$\Delta{}E_{\ab{sol}}$
\\ \hline
 this work&-2423.0&-2421.3&-7.5&5.8&-28.6&-22.8\\
 \amber{}&-2428.3&-2433.0&-17.6&22.4&-27.7&-5.3\\
 expt. \cite{catecholexpt}&--&--&-9.3&--&--&--\\
\end{tabular}
\end{table}


Conventional \abinitio{} calculations are typically limited to only a few hundred atoms at most. 
However, with recent advances in linear-scaling density functional theory (LS-DFT) 
approaches \cite{Goedecker_review} a number of codes \cite{SKYLARIS2005, BOWLER2006, artacho2008} 
have been developed which are capable of performing calculations
with many thousands of atoms. The combination of LS-DFT with implicit solvent models
would enable higly realistic simulations of important phenomena such as biomolecular processes or the chemical modification
and self-assembly of nanostructures.

\begin{figure}[!htbp]
\centering
\includegraphics[viewport=20 10 910 595,clip,width=\figwidth]{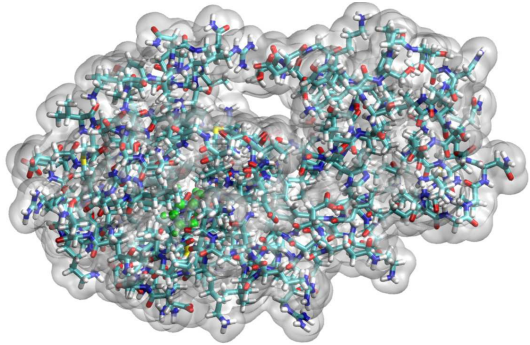}
\caption{
\label{fig:lysozyme} 
An overview of the lysozyme-catechol complex, with the dielectric cavity
indicated in grey and catechol in green.}
\end{figure}

As a demonstration of the potential applications of this approach in large-scale DFT calculations, 
we have implemented our solvent model in the LS-DFT code \onetep{} \cite{SKYLARIS2005} and used it to calculate the free energy of solvation for a 2615-atom system composed of the catechol ligand bound to a L99A/M102Q mutant of the T4 lysozyme. The results of this calculation are shown in~table~\ref{tab:lysozyme}
and the overview of the system in question, along with an outline of the dielectric cavity (at
$\rho=\rho_0$) is shown in fig.~\ref{fig:lysozyme}. 
\amber{} greatly overestimates the solvation energy of catechol, and consequently the binding energy in solvent differs significantly between the two models. The need for an \abinitio{} model, where the density is polarized by the solvent is demonstrated by the different behavior of $\Delta{}G_{\ab{host}}$ (as compared to $\Delta{}G_{\ab{cplx}}$) between the proposed model and \amber{}.

In summary, we have outlined and validated an implicit solvent model for \abinitio{} calculations, which, despite using only two parameters, offers a substantial improvement over existing models, as measured
by the agreement of absolute and relative free energies of solvation with experiment (compared to PCM and \amber{}) or the number of parameters needed to achieve similar agreement (compared to SMD). We have shown how the implementation of the proposed model in the LS-DFT code \onetep{} paves the way for first-principles implicit-solvent calculations for molecules with thousands of atoms.

\acknowledgments
The calculations were carried out on the Iridis3 supercomputer of the University of Southampton \revision{and on the Darwin Supercomputer of the University of Cambridge High Performance
Computing Service provided by Dell Inc. using Strategic Research Infrastructure Funding from
the Higher Education Funding Council for England}.
JD, CKS, AAM and MCP acknowledge funding from EPSRC Grant EP/G055904/1. HHH acknowledges the support of the Yousef Jameel PhD scholarship adminstered by the University of Cambridge. CKS would like to thank the Royal Society for a University Research Fellowship. AAM acknowledges the support of an RCUK academic fellowship. MCP acknowledges funding from EPSRC Grant EP/F0327731/1. 

\nocite{*}

\bibliographystyle{aipnum4-1}

%


\begin{thebibliography}{26}%
\makeatletter
\providecommand \@ifxundefined [1]{%
 \@ifx{#1\undefined}
}%
\providecommand \@ifnum [1]{%
 \ifnum #1\expandafter \@firstoftwo
 \else \expandafter \@secondoftwo
 \fi
}%
\providecommand \@ifx [1]{%
 \ifx #1\expandafter \@firstoftwo
 \else \expandafter \@secondoftwo
 \fi
}%
\providecommand \natexlab [1]{#1}%
\providecommand \enquote  [1]{``#1''}%
\providecommand \bibnamefont  [1]{#1}%
\providecommand \bibfnamefont [1]{#1}%
\providecommand \citenamefont [1]{#1}%
\providecommand \href@noop [0]{\@secondoftwo}%
\providecommand \href [0]{\begingroup \@sanitize@url \@href}%
\providecommand \@href[1]{\@@startlink{#1}\@@href}%
\providecommand \@@href[1]{\endgroup#1\@@endlink}%
\providecommand \@sanitize@url [0]{\catcode `\\12\catcode `\$12\catcode
  `\&12\catcode `\#12\catcode `\^12\catcode `\_12\catcode `\%12\relax}%
\providecommand \@@startlink[1]{}%
\providecommand \@@endlink[0]{}%
\providecommand \url  [0]{\begingroup\@sanitize@url \@url }%
\providecommand \@url [1]{\endgroup\@href {#1}{\urlprefix }}%
\providecommand \urlprefix  [0]{URL }%
\providecommand \Eprint [0]{\href }%
\providecommand \doibase [0]{http://dx.doi.org/}%
\providecommand \selectlanguage [0]{\@gobble}%
\providecommand \bibinfo  [0]{\@secondoftwo}%
\providecommand \bibfield  [0]{\@secondoftwo}%
\providecommand \translation [1]{[#1]}%
\providecommand \BibitemOpen [0]{}%
\providecommand \bibitemStop [0]{}%
\providecommand \bibitemNoStop [0]{.\EOS\space}%
\providecommand \EOS [0]{\spacefactor3000\relax}%
\providecommand \BibitemShut  [1]{\csname bibitem#1\endcsname}%
\let\auto@bib@innerbib\@empty
\bibitem [{\citenamefont {Henchman}\ and\ \citenamefont
  {Essex}(1999)}]{henchman1999}%
  \BibitemOpen
  \bibfield  {author} {\bibinfo {author} {\bibfnamefont {R.~H.}\ \bibnamefont
  {Henchman}}\ and\ \bibinfo {author} {\bibfnamefont {J.~W.}\ \bibnamefont
  {Essex}},\ }\href@noop {} {\bibfield  {journal} {\bibinfo  {journal} {J.
  Comp. Chem.}\ }\textbf {\bibinfo {volume} {20}},\ \bibinfo {pages} {499}
  (\bibinfo {year} {1999})}\BibitemShut {NoStop}%
\bibitem [{\citenamefont {Tomasi}, \citenamefont {Mennucci},\ and\
  \citenamefont {Cammi}(2005)}]{Tomasi2}%
  \BibitemOpen
  \bibfield  {author} {\bibinfo {author} {\bibfnamefont {J.}~\bibnamefont
  {Tomasi}}, \bibinfo {author} {\bibfnamefont {B.}~\bibnamefont {Mennucci}}, \
  and\ \bibinfo {author} {\bibfnamefont {R.}~\bibnamefont {Cammi}},\ }\href
  {\doibase 10.1021/cr9904009} {\bibfield  {journal} {\bibinfo  {journal}
  {Chem. Rev.}\ }\textbf {\bibinfo {volume} {105}},\ \bibinfo {pages} {2999}
  (\bibinfo {year} {2005})}\BibitemShut {NoStop}%
\bibitem [{\citenamefont {Tomasi}\ and\ \citenamefont
  {Persico}(1994)}]{Tomasi1}%
  \BibitemOpen
  \bibfield  {author} {\bibinfo {author} {\bibfnamefont {J.}~\bibnamefont
  {Tomasi}}\ and\ \bibinfo {author} {\bibfnamefont {M.}~\bibnamefont
  {Persico}},\ }\href {\doibase 10.1021/cr00031a013} {\bibfield  {journal}
  {\bibinfo  {journal} {Chem. Rev.}\ }\textbf {\bibinfo {volume} {94}},\
  \bibinfo {pages} {2027} (\bibinfo {year} {1994})}\BibitemShut {NoStop}%
\bibitem [{\citenamefont {Klamt}\ and\ \citenamefont
  {Sch\"u\"urmann}(1993)}]{Klamt1993}%
  \BibitemOpen
  \bibfield  {author} {\bibinfo {author} {\bibfnamefont {A.}~\bibnamefont
  {Klamt}}\ and\ \bibinfo {author} {\bibnamefont {Sch\"u\"urmann}},\
  }\href@noop {} {\bibfield  {journal} {\bibinfo  {journal} {J. Chem. Soc.
  Perkin Trans.}\ }\textbf {\bibinfo {volume} {2}},\ \bibinfo {pages} {799}
  (\bibinfo {year} {1993})}\BibitemShut {NoStop}%
\bibitem [{\citenamefont {Fattebert}\ and\ \citenamefont
  {Gygi}(2002)}]{FattebertGygi}%
  \BibitemOpen
  \bibfield  {author} {\bibinfo {author} {\bibfnamefont {J.-L.}\ \bibnamefont
  {Fattebert}}\ and\ \bibinfo {author} {\bibfnamefont {F.}~\bibnamefont
  {Gygi}},\ }\href {\doibase 10.1002/jcc.10069} {\bibfield  {journal} {\bibinfo
   {journal} {Journal of Computational Chemistry}\ }\textbf {\bibinfo {volume}
  {23}},\ \bibinfo {pages} {662} (\bibinfo {year} {2002})}\BibitemShut
  {NoStop}%
\bibitem [{\citenamefont {Fattebert}\ and\ \citenamefont
  {Gygi}(2003)}]{FattebertGygi2003}%
  \BibitemOpen
  \bibfield  {author} {\bibinfo {author} {\bibfnamefont {J.-L.}\ \bibnamefont
  {Fattebert}}\ and\ \bibinfo {author} {\bibfnamefont {F.}~\bibnamefont
  {Gygi}},\ }\href {\doibase 10.1002/qua.10548} {\bibfield  {journal} {\bibinfo
   {journal} {Int. J. Quantum Chem.}\ }\textbf {\bibinfo {volume} {93}},\
  \bibinfo {pages} {139} (\bibinfo {year} {2003})}\BibitemShut {NoStop}%
\bibitem [{\citenamefont {Scherlis}\ \emph {et~al.}(2006)\citenamefont
  {Scherlis}, \citenamefont {Fattebert}, \citenamefont {Gygi}, \citenamefont
  {Cococcioni},\ and\ \citenamefont {Marzari}}]{Scherlis}%
  \BibitemOpen
  \bibfield  {author} {\bibinfo {author} {\bibfnamefont {D.}~\bibnamefont
  {Scherlis}}, \bibinfo {author} {\bibfnamefont {J.}~\bibnamefont {Fattebert}},
  \bibinfo {author} {\bibfnamefont {F.}~\bibnamefont {Gygi}}, \bibinfo {author}
  {\bibfnamefont {M.}~\bibnamefont {Cococcioni}}, \ and\ \bibinfo {author}
  {\bibfnamefont {N.}~\bibnamefont {Marzari}},\ }\href@noop {} {\bibfield
  {journal} {\bibinfo  {journal} {J. Chem. Phys.}\ }\textbf {\bibinfo {volume}
  {124}},\ \bibinfo {pages} {{074103}} (\bibinfo {year} {2006})}\BibitemShut
  {NoStop}%
\bibitem [{\citenamefont {Marenich}\ \emph {et~al.}(2009)\citenamefont
  {Marenich} \emph {et~al.}}]{soldb}%
  \BibitemOpen
  \bibfield  {author} {\bibinfo {author} {\bibfnamefont {A.}~\bibnamefont
  {Marenich}} \emph {et~al.},\ }\href@noop {} {\enquote {\bibinfo {title}
  {Minnesota solvation database, version 2009},}\ }\bibinfo {howpublished}
  {University of Minnesota, Minneapolis} (\bibinfo {year} {2009})\BibitemShut
  {NoStop}%
\bibitem [{\citenamefont {Makov}\ and\ \citenamefont
  {Payne}(1995)}]{MakovPayne}%
  \BibitemOpen
  \bibfield  {author} {\bibinfo {author} {\bibfnamefont {G.}~\bibnamefont
  {Makov}}\ and\ \bibinfo {author} {\bibfnamefont {M.~C.}\ \bibnamefont
  {Payne}},\ }\href {\doibase 10.1103/PhysRevB.51.4014} {\bibfield  {journal}
  {\bibinfo  {journal} {Physical Review B}\ }\textbf {\bibinfo {volume} {51}},\
  \bibinfo {pages} {4014} (\bibinfo {year} {1995})}\BibitemShut {NoStop}%
\bibitem [{\citenamefont {Case}\ \emph {et~al.}(2008)\citenamefont {Case} \emph
  {et~al.}}]{amber}%
  \BibitemOpen
  \bibfield  {author} {\bibinfo {author} {\bibfnamefont {D.~A.}\ \bibnamefont
  {Case}} \emph {et~al.},\ }\href {http://amber.scripps.edu/\#Amber10} {\emph
  {\bibinfo {title} {Amber 10}}}\ (\bibinfo  {publisher} {University of
  California},\ \bibinfo {address} {San Francisco},\ \bibinfo {year}
  {2008})\BibitemShut {NoStop}%
\bibitem [{\citenamefont {Beck}(2000)}]{multigrid}%
  \BibitemOpen
  \bibfield  {author} {\bibinfo {author} {\bibfnamefont {T.~L.}\ \bibnamefont
  {Beck}},\ }\href {\doibase 10.1103/RevModPhys.72.1041} {\bibfield  {journal}
  {\bibinfo  {journal} {Rev. Mod. Phys.}\ }\textbf {\bibinfo {volume} {72}},\
  \bibinfo {pages} {1041} (\bibinfo {year} {2000})}\BibitemShut {NoStop}%
\bibitem [{\citenamefont {Holst}\ and\ \citenamefont {Saied}(1993)}]{Holst}%
  \BibitemOpen
  \bibfield  {author} {\bibinfo {author} {\bibfnamefont {M.}~\bibnamefont
  {Holst}}\ and\ \bibinfo {author} {\bibfnamefont {F.}~\bibnamefont {Saied}},\
  }\href@noop {} {\bibfield  {journal} {\bibinfo  {journal} {J. Comput. Chem}\
  }\textbf {\bibinfo {volume} {14}},\ \bibinfo {pages} {105} (\bibinfo {year}
  {1993})}\BibitemShut {NoStop}%
\bibitem [{\citenamefont {Trottenberg}, \citenamefont {Oosterle},\ and\
  \citenamefont {Schuller}(2000)}]{defcorr}%
  \BibitemOpen
  \bibfield  {author} {\bibinfo {author} {\bibfnamefont {U.}~\bibnamefont
  {Trottenberg}}, \bibinfo {author} {\bibfnamefont {C.~W.}\ \bibnamefont
  {Oosterle}}, \ and\ \bibinfo {author} {\bibfnamefont {A.}~\bibnamefont
  {Schuller}},\ }\href@noop {} {\emph {\bibinfo {title} {Multigrid}}}\
  (\bibinfo  {publisher} {Academic Press},\ \bibinfo {year} {2000})\BibitemShut
  {NoStop}%
\bibitem [{\citenamefont {Mobley}\ \emph {et~al.}(2008)\citenamefont {Mobley},
  \citenamefont {Barber}, \citenamefont {Fennell},\ and\ \citenamefont
  {Dill}}]{Mobley_asymmetry}%
  \BibitemOpen
  \bibfield  {author} {\bibinfo {author} {\bibfnamefont {D.~L.}\ \bibnamefont
  {Mobley}}, \bibinfo {author} {\bibfnamefont {A.~E.}\ \bibnamefont {Barber}},
  \bibinfo {author} {\bibfnamefont {C.~J.}\ \bibnamefont {Fennell}}, \ and\
  \bibinfo {author} {\bibfnamefont {K.~A.}\ \bibnamefont {Dill}},\ }\href
  {\doibase 10.1021/jp709958f} {\bibfield  {journal} {\bibinfo  {journal} {J.
  Phys. Chem. B}\ }\textbf {\bibinfo {volume} {112}},\ \bibinfo {pages} {2405}
  (\bibinfo {year} {2008})}\BibitemShut {NoStop}%
\bibitem [{\citenamefont {Zhao}, \citenamefont {Schultz},\ and\ \citenamefont
  {Truhlar}(2006)}]{M052X}%
  \BibitemOpen
  \bibfield  {author} {\bibinfo {author} {\bibfnamefont {Y.}~\bibnamefont
  {Zhao}}, \bibinfo {author} {\bibfnamefont {N.~E.}\ \bibnamefont {Schultz}}, \
  and\ \bibinfo {author} {\bibfnamefont {D.~G.}\ \bibnamefont {Truhlar}},\
  }\href {\doibase 10.1021/ct0502763} {\bibfield  {journal} {\bibinfo
  {journal} {J. Chem. Theor. Comput.}\ }\textbf {\bibinfo {volume} {2}},\
  \bibinfo {pages} {364} (\bibinfo {year} {2006})}\BibitemShut {NoStop}%
\bibitem [{\citenamefont {Floris}, \citenamefont {Tomasi},\ and\ \citenamefont
  {Ahuir}(1991)}]{FlorisTomasi}%
  \BibitemOpen
  \bibfield  {author} {\bibinfo {author} {\bibfnamefont {F.~M.}\ \bibnamefont
  {Floris}}, \bibinfo {author} {\bibfnamefont {J.}~\bibnamefont {Tomasi}}, \
  and\ \bibinfo {author} {\bibfnamefont {J.~L.~P.}\ \bibnamefont {Ahuir}},\
  }\href {\doibase 10.1002/jcc.540120703} {\bibfield  {journal} {\bibinfo
  {journal} {J. Comput. Chem.}\ }\textbf {\bibinfo {volume} {12}},\ \bibinfo
  {pages} {784} (\bibinfo {year} {1991})}\BibitemShut {NoStop}%
\bibitem [{Note1()}]{Note1}%
  \BibitemOpen
  \bibinfo {note} {SMD is a recently proposed model based on the
  integral-equation-formalism PCM (IEF-PCM), which yields excellent agreement
  with experiment. This requires, however, making use of an extensive set of
  parameters to describe the solute (intrinsic Coulomb radii, atomic surface
  tension parameters) and the solvent (refractive index and acidity and
  basicity parameters; in addition to the dielectric constant and bulk surface
  tension needed in the proposed model).}\BibitemShut {Stop}%
\bibitem [{\citenamefont {Marenich}, \citenamefont {Cramer},\ and\
  \citenamefont {Truhlar}(2009)}]{SMD}%
  \BibitemOpen
  \bibfield  {author} {\bibinfo {author} {\bibfnamefont {A.~V.}\ \bibnamefont
  {Marenich}}, \bibinfo {author} {\bibfnamefont {C.~J.}\ \bibnamefont
  {Cramer}}, \ and\ \bibinfo {author} {\bibfnamefont {D.~G.}\ \bibnamefont
  {Truhlar}},\ }\href@noop {} {\bibfield  {journal} {\bibinfo  {journal}
  {Journal of Chemical Theory and Computation}\ }\textbf {\bibinfo {volume}
  {5}},\ \bibinfo {pages} {2447} (\bibinfo {year} {2009})}\BibitemShut
  {NoStop}%
\bibitem [{\citenamefont {Sanchez}, \citenamefont {Sued},\ and\ \citenamefont
  {Scherlis}(2009)}]{Scherlis2009}%
  \BibitemOpen
  \bibfield  {author} {\bibinfo {author} {\bibfnamefont {V.}~\bibnamefont
  {Sanchez}}, \bibinfo {author} {\bibfnamefont {M.}~\bibnamefont {Sued}}, \
  and\ \bibinfo {author} {\bibfnamefont {D.}~\bibnamefont {Scherlis}},\
  }\href@noop {} {\bibfield  {journal} {\bibinfo  {journal} {J Chem Phys}\
  }\textbf {\bibinfo {volume} {131}},\ \bibinfo {pages} {174108} (\bibinfo
  {year} {2009})}\BibitemShut {NoStop}%
\bibitem [{\citenamefont {Nicholls}\ \emph {et~al.}(2008)\citenamefont
  {Nicholls} \emph {et~al.}}]{blind1}%
  \BibitemOpen
  \bibfield  {author} {\bibinfo {author} {\bibfnamefont {A.}~\bibnamefont
  {Nicholls}} \emph {et~al.},\ }\href {\doibase 10.1021/jm070549+} {\bibfield
  {journal} {\bibinfo  {journal} {J. Med. Chem.}\ }\textbf {\bibinfo {volume}
  {51}},\ \bibinfo {pages} {769} (\bibinfo {year} {2008})}\BibitemShut
  {NoStop}%
\bibitem [{\citenamefont {Guthrie}(2009)}]{blind2}%
  \BibitemOpen
  \bibfield  {author} {\bibinfo {author} {\bibfnamefont {J.~P.}\ \bibnamefont
  {Guthrie}},\ }\href {\doibase 10.1021/jp806724u} {\bibfield  {journal}
  {\bibinfo  {journal} {J. Phys. Chem. B}\ }\textbf {\bibinfo {volume} {113}},\
  \bibinfo {pages} {4501} (\bibinfo {year} {2009})}\BibitemShut {NoStop}%
\bibitem [{\citenamefont {Mordasini}\ and\ \citenamefont
  {McCammon}(2000)}]{catecholexpt}%
  \BibitemOpen
  \bibfield  {author} {\bibinfo {author} {\bibfnamefont {T.~Z.}\ \bibnamefont
  {Mordasini}}\ and\ \bibinfo {author} {\bibfnamefont {J.~A.}\ \bibnamefont
  {McCammon}},\ }\href@noop {} {\bibfield  {journal} {\bibinfo  {journal} {J.
  Phys. Chem. B}\ }\textbf {\bibinfo {volume} {104}},\ \bibinfo {pages} {360}
  (\bibinfo {year} {2000})}\BibitemShut {NoStop}%
\bibitem [{\citenamefont {Goedecker}(1999)}]{Goedecker_review}%
  \BibitemOpen
  \bibfield  {author} {\bibinfo {author} {\bibfnamefont {S.}~\bibnamefont
  {Goedecker}},\ }\href@noop {} {\bibfield  {journal} {\bibinfo  {journal}
  {Rev. Mod. Phys.}\ }\textbf {\bibinfo {volume} {71}},\ \bibinfo {pages}
  {1085} (\bibinfo {year} {1999})}\BibitemShut {NoStop}%
\bibitem [{\citenamefont {Skylaris}\ \emph {et~al.}(2005)\citenamefont
  {Skylaris}, \citenamefont {Haynes}, \citenamefont {Mostofi},\ and\
  \citenamefont {Payne}}]{SKYLARIS2005}%
  \BibitemOpen
  \bibfield  {author} {\bibinfo {author} {\bibfnamefont {C.-K.}\ \bibnamefont
  {Skylaris}}, \bibinfo {author} {\bibfnamefont {P.~D.}\ \bibnamefont
  {Haynes}}, \bibinfo {author} {\bibfnamefont {A.~A.}\ \bibnamefont {Mostofi}},
  \ and\ \bibinfo {author} {\bibfnamefont {M.~C.}\ \bibnamefont {Payne}},\
  }\href@noop {} {\bibfield  {journal} {\bibinfo  {journal} {J. Chem. Phys.}\
  }\textbf {\bibinfo {volume} {122}},\ \bibinfo {pages} {084119} (\bibinfo
  {year} {2005})}\BibitemShut {NoStop}%
\bibitem [{\citenamefont {Bowler}\ \emph {et~al.}(2006)\citenamefont {Bowler},
  \citenamefont {Choudhury}, \citenamefont {Gillan},\ and\ \citenamefont
  {Miyazaki}}]{BOWLER2006}%
  \BibitemOpen
  \bibfield  {author} {\bibinfo {author} {\bibfnamefont {D.~R.}\ \bibnamefont
  {Bowler}}, \bibinfo {author} {\bibfnamefont {R.}~\bibnamefont {Choudhury}},
  \bibinfo {author} {\bibfnamefont {M.~J.}\ \bibnamefont {Gillan}}, \ and\
  \bibinfo {author} {\bibfnamefont {T.}~\bibnamefont {Miyazaki}},\ }\href@noop
  {} {\bibfield  {journal} {\bibinfo  {journal} {phys. stat. sol. (b)}\
  }\textbf {\bibinfo {volume} {243}},\ \bibinfo {pages} {989} (\bibinfo {year}
  {2006})}\BibitemShut {NoStop}%
\bibitem [{\citenamefont {Artacho}\ \emph {et~al.}(2008)\citenamefont {Artacho}
  \emph {et~al.}}]{artacho2008}%
  \BibitemOpen
  \bibfield  {author} {\bibinfo {author} {\bibfnamefont {E.}~\bibnamefont
  {Artacho}} \emph {et~al.},\ }\href@noop {} {\bibfield  {journal} {\bibinfo
  {journal} {J. Phys. Cond. Mat.}\ }\textbf {\bibinfo {volume} {20}},\ \bibinfo
  {pages} {064208} (\bibinfo {year} {2008})}\BibitemShut {NoStop}%
\end{thebibliography}

\end{document}